**Powering Monolithic and Hybrid Organic Optical Waveguides via Integrated Focused Micro-LEDs for Sustainable Photonic Circuits**

*Ankur Khapre,[#] Avulu Vinod Kumar,[#] and Rajadurai Chandrasekar\**

Advanced Photonic Materials and Technology Laboratory

School of Chemistry and Centre for Nanotechnology, University of Hyderabad, Prof. C. R. Rao Road, Gachibowli, Hyderabad 500046, Telangana, India

E-mail: *r.chandrasekar@uohyd.ac.in*

Keywords: (organic flexible crystals, electrically pumped waveguides, visible light photonics, hybrid waveguides)

[#]Equal contribution of authors

In the domain of mechanophotonics, achieving real-time applicability of organic crystals in visible light communication (VLC) technologies necessitates affordable light-emitting diodes (LEDs) as sources of light to run photonic devices through sustainable methods. Here in, we demonstrate an efficient strategy to excite (*Z*)-3-(3',5'-bis(trifluoromethyl)-[1,1'-biphenyl]-4-yl)-2-(4-methoxyphenyl) acrylonitrile ($CF_3OMe$), 9,10-bis(phenylethynyl)anthracene (BPEA) and 2,2'-((1*E*,1'*E*)-hydrazine-1,2-diylidenebis(methaneylylidene))diphenol (SAA) flexible crystal waveguides utilizing UV LED source and transduce respective blue, orange and yellow fluorescence signals. The capability of the focused LED lies in its ability to (i) energize mechanically bent crystals at an angle of 180°, (ii) evanescently excite the FL of a SAA waveguide using the FL of $CF_3OMe$ waveguide through energy transfer, and (iii) excite and split different signals in a 2×2 hybrid directional coupler based on SSA-BPEA crystals. These demonstrations underscore the practicality of the proposed technique for sustainable applications in photonic systems related to VLC.

**Introduction**

Visible Light Communication (VLC) technology emerges as a promising solution for sustainable and environmentally friendly technologies by harnessing abundant natural light.[1] The research interest in VLC has surged significantly owing to its potential applications in Light Fidelity (Li-Fi) technology, indoor devices, portable communication devices, under water Environments and multi-user-compatible technologies.[2-6] VLC's appeal lies in its ability to utilize ambient light sources, making it particularly suitable for both indoor and outdoor settings, be it sunlight or light-emitting diodes (LEDs). To facilitate the widespread adoption of VLC, there is a critical need for the development of innovative devices that can effectively operate with the ambient light commonly found in human environments. In this context, the utilization of commercial LEDs as optical sources becomes paramount for the success of VLC systems. Therefore, the establishment of inventive device-powering strategies, with a focus on leveraging commercial LEDs, becomes a crucial aspect of VLC technology. These strategies not only contribute to the efficiency and reliability of VLC devices but also align with the broader goal of sustainable and energy-efficient technologies.

Recent breakthroughs in flexible organic crystals have led to the development of photonics devices such as waveguides,[7-10] resonators,[14-17] directional couplers,[18] and integrated circuits[16-19] powered by commercial lasers. However, lasers have drawbacks, including cost, power consumption, and material damage. Transitioning to commercial LEDs as light sources presents a cost-effective alternative. Ingenious design strategies, incorporating the right materials and focused LED sources, are crucial for realizing functional photonic systems and circuits, offering a practical solution to the limitations associated with using lasers.

Hence, focused LED-powered photonic devices, showcasing both active (photoluminescence-guided) and passive (same input light-guided) waveguides, cavities, and hybrid photonic circuits, represent an initial step toward utilizing low-power sources in photonic systems. To meet the demands of VLC technologies, we explore the use of a cost-effective commercial UV LED. We developed focused LED-powered organic waveguides and circuits, operating in a complete visible spectral bandwidth, using blue-emissive ($Z$)-3-(3',5'-bis(trifluoromethyl)-[1,1'-biphenyl]-4-yl)-2-(4-methoxyphenyl) acrylonitrile ($CF_3OMe$), orange-emissive 9,10-bis(phenylethynyl)anthracene (BPEA), and yellow-emissive 2,2'-((1$E$,1'$E$)-hydrazine-1,2-diylidenebis(methaneylylidene))diphenol (SAA) flexible crystals. For constructing the LED-powered photonic circuit technology, a glass slide is used, with a thin aluminum foil inscribed with a 40 μm diameter hole on the rear side to facilitate light illumination. A UV LED is strategically positioned on the foil, aligning with the hole to enable precise focused light transmission. The light emanating from the aperture powers the organic crystal optical waveguides and 2×2 hybrid directional coupler, strategically placed on the opposite side of the glass slide to receive the emitted LED light. This proof-of-principle device efficiently operates when powered with a UV LED, showcasing active and passive light-guiding capabilities based on reabsorption and energy transfer principles.

## Results and Discussion

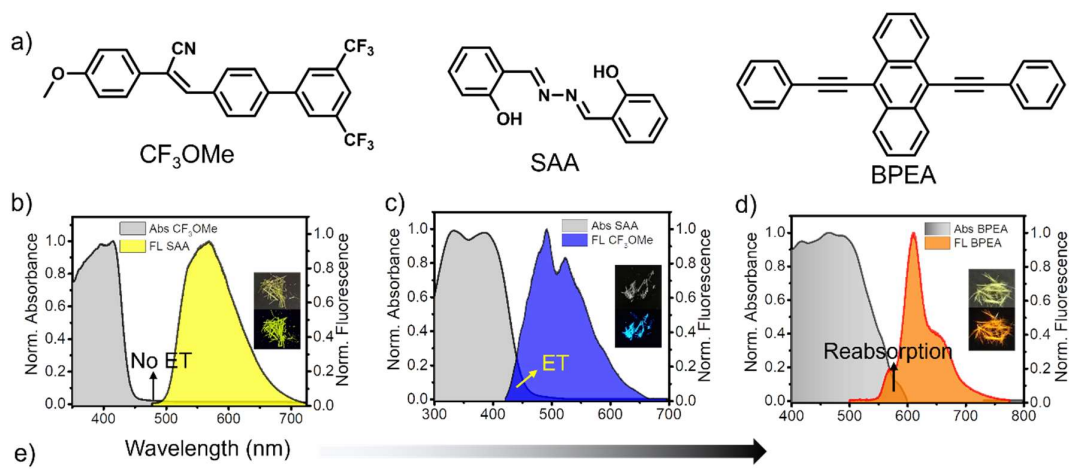

**Figure 1.** a) Molecular structure of CF₃OMe, SAA and BPEA. b) Solid-state absorption of CF₃OMe and FL of SAA. c) Solid-state absorption of SAA and FL of CF₃OMe. d) Solid state absorption and emission of BPPEA. e) Details about the material's absorption maxima, emission bandwidth, photoluminescence quantum yield, reabsorption and energy transfer.

| Material | Absorption (nm) | Emission BW (nm) | PLQY (%) | Reabsorbed BW (nm) | Energy Transfer |
|---|---|---|---|---|---|
| CF₃OMe | Tail ≈ 470 | $\lambda_1 \approx 420 - 665$<br>$\lambda_{max} \approx 492/524$ | 36 | $-\lambda_1 \approx 435 - 665$ | $\lambda_2$ (ET$_A$) ≈ 420 – 500 nm (CF₃OMe → SAA) |
| SAA | Tail ≈ 500 | $\lambda_2 \approx 495 - 720$<br>$\lambda_{max} \approx 568$ | 23 | $-\lambda_2 \approx 515 - 720$ | $\lambda_3$ (ET$_C$) ≈ 495 – 600 nm (SAA → BPEA) |
| BPEA | Tail ≈ 600 | $\lambda_3 \approx 550 - 770$<br>$\lambda_{max} \approx 595$ | 24 | $-\lambda_3 \approx 580 - 770$ | $\lambda_3$ (ET$_B$) ≈ 420 – 600 nm (CF₃OMe → BPEA) |

The desired molecules shown in Figure 1a were synthesized according to the procedures given in the supporting information.[17,23,24] The absorption of CF₃OMe crystal expands from the UV region to the blue part of the visible spectrum (absorption tail ≈ 450 nm, Figure 1b). The blue FL of CF₃OMe covers $\lambda_1$ ≈420–665 nm with a $\lambda_{max}$ at 475 nm and PLQY of ≈36%. Similarly, the absorption of the SAA compound extended up to 500 nm, while its emission ranged between $\lambda_2$ ≈495–720 nm (PLQY ≈23%). Importantly, the absorption of SAA and emission of CF₃OMe share a common region from 420–465 nm, giving rise to the possibility of energy transfer from CF₃OMe to SAA crystal. The absorption of BPEA extended to 600 nm and its orange emission covered from $\lambda_3$ ≈550-770 nm (PLQY ≈24%) with a $\lambda_{max}$ at 595 nm.

Compound SAA readily crystallized as thin acicular crystals from ethanol:dichloromethane (3:1) solution. The single crystal X-ray diffraction analysis suggested SAA crystallized in the monoclinic system. The molecule comprised a near planar geometry in the crystal packing imparted by intramolecular hydrogen bonding (O-H⋯N =1.848 Å) (Figure 2a). The interlayer stacking distance between adjacent molecules is ≈3.464 Å (Figure 2b). Further, the molecular packing adopted a crisscross-type arrangement typically observed in many flexible Schiff base derivatives (Figure 2c). The molecular packing along (10$\bar{1}$) and (010) are depicted in Figure 2d-f, respectively. Possibly, the intermolecular interactions (C-O⋯H =2.438 Å) and crisscross arrangement impart mechanical flexibility in SAA crystals. To verify the same, a milli-meter-sized SAA crystal was subjected to external mechanical stress.

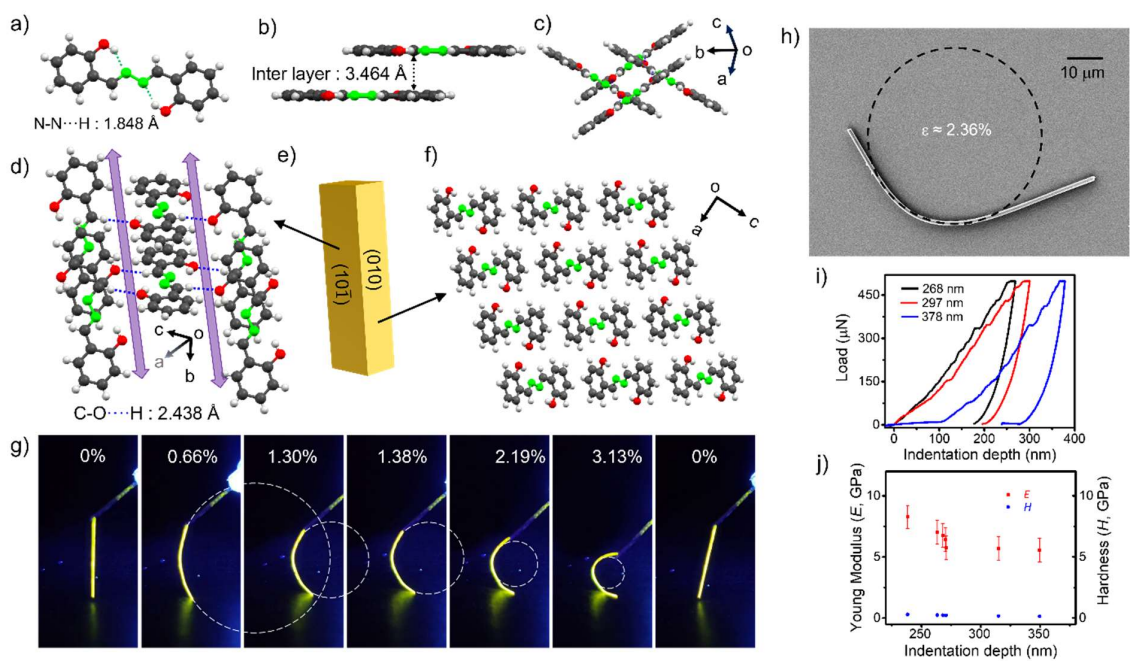

**Figure 2.** a) Single crystal X-ray structure of SAA. The green dashed line represents the intramolecular N⋯H-O hydrogen bond. b) Side view of the SAA structure and dashed line arrow denotes intermolecular distance. c) Molecular crisscross-packed structure visualized down the (10$\bar{1}$) plane. d) The molecular packing visualized along the (10$\bar{1}$) plane (Blue dashed line represents the intermolecular hydrogen bond between C-O⋯H). e) Cartoon shows the planes of the SAA single crystal. f) Molecular packing visualized down the (010) plane along the *b*-axis. g) Photographs of a thin SAA needle-shaped molecular crystal exhibiting elastic mechanical flexibility, captured under UV torch illumination at different strain values. h) FESEM image of a SAA microcrystal. i) The plot between load and indentation depth attained during nanoindentation performed on the thicker (10$\bar{1}$) face of SAA crystal. Black, red and blue lines correspond to different contact depths, 268, 297 and 378 nm, respectively. j) The corresponding Young's modulus (E) and Hardness (H) values for the same crystal.

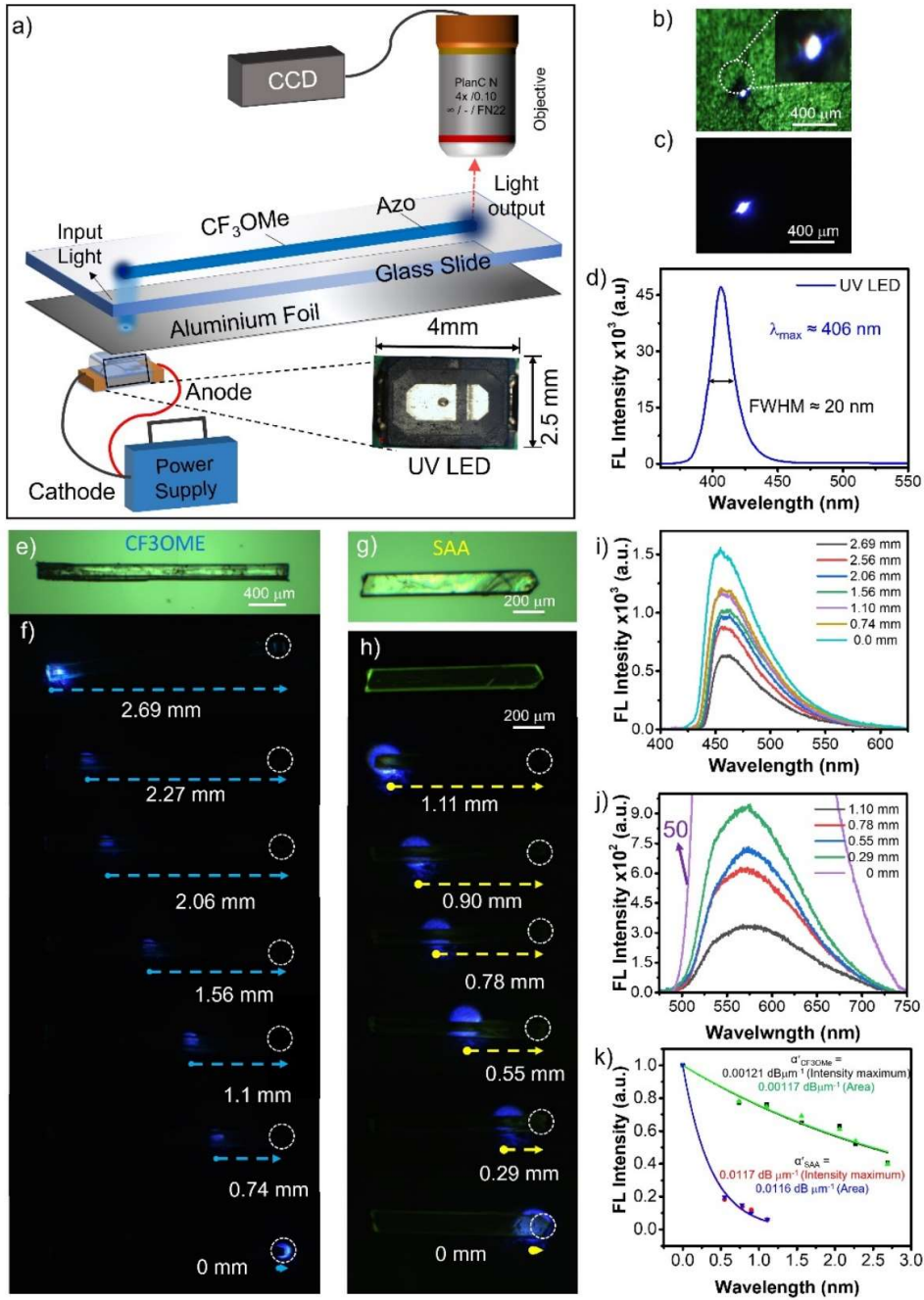

**Figure 3.** a) Graphical representation of the device used for the demonstration of a focused UV LED-powered orgnaic crystal waveguide. (The bottom right inset shows the photograph of a commercial UV LED). b-c) Confocal optical and FL images of UV LED light emanating out of 40 μm diameter hole in aluminum foil, respectively. d) The optical emission spectrum of UV LED showing λ$_{max}$ ≈406 nm and FWHM≈20 nm. e,f) Confocal optical and FL images of CF$_3$OMe, respectively. g,h) Confocal optical and FL images of SAA, respectively. i,j) FL spectra of CF$_3$OMe and SAA, respectively. k) I$_{tip}$/I$_{body}$ vs distance plot for CF$_3$OMe and SAA (The optical loss, α′$_{A\ was}$ calculated using emission intensity maxima and α′$_B$ calculated using integrated FL spectral area under the curve).

The strain on the crystal gradually increased from 0.66% to 3.13% and slowly released the strain to confirm elastic behavior (Figure 2g, Supporting Video 1). The reversible mechanical response could be seen multiple times without considerable damage to the crystal. The smooth surface morphology of a bent SAA crystal suggested no physical damage upon bending (Figure 2h). The quantitative mechanical property investigations were carried out by subjecting the SAA crystal to nanoindentation. The load versus displacement curve obtained on the widest crystal plane exhibited a large residual depth, indicating the elastic mechanical behavior of SAA crystals (Figure 2i). The Young's modulus ($E$) and hardness ($H$) of the crystal were measured to be 0.22 GPa and 6.49 GPa, respectively (Figure 2j).

To assess the viability of employing LED light sources for photonic applications, a commercial micro-LED light source with a peak wavelength of approximately 406 nm (covering the 350-475 nm range) and a narrow full-width at half-maximum (FWHM) of around 20 nm was chosen (Figure 3a-d). An aluminum foil with a 40 µm diameter hole was positioned at the bottom of a coverslip. The LED was placed below the hole, and the power emerging from the hole to the opposite side of the coverslip was approximately 37 µW. A $CF_3OMe$ single crystal (length, $L\approx2.69$ mm, and thickness, $T\approx143$ µm) was placed atop the coverslip, and its left tip was precisely aligned with the emanating focused LED light (Figure 3e,f). When the LED was turned on, the $CF_3OMe$ crystal emitted blue FL ($\lambda_1 \approx 420–665$ nm), transducing the FL to the opposite end confirming its active-type waveguiding capability (Figure 3i). To assess the optical loss coefficient ($\alpha'$) of the crystal waveguide, it was manipulated with a needle, exposing different zones along its long axis to LED light, and recording the resulting FL at the right end of the waveguide (Figure 3e-f,i). The distances between excitation and guided FL collection tip in the waveguide were 2.69, 2.57, 2.06, 1.56, 1.1, 0.74, and 0 mm. As anticipated, the FL intensity rapidly increased as the LED source approached closer to the collection point. The $\alpha'$ was determined to be 0.00121 dB µm$^{-1}$ (Figure 3k).

Demonstrating the broad applicability of LED light as an optical source for powering waveguides necessitates showcasing light coupling into waveguides made of diverse crystalline materials. Hence, similar to the aforementioned experiment, an SAA crystal's ($L\approx1.11$ mm) left tip was optically excited with focused LED light. At the excitation point, the generated yellow FL ($\lambda_2\approx495–720$ nm) transduced to the opposite tip of the crystal (515–720 8nm) with a slight reduction in bandwidth due to the reabsorption of the self-propagating optical signal (Figure 3g,j). The FL of the SAA crystal was recorded at the right end by varying the distance between the LED-light source and the FL collection point (0, 0.29, 0.55, 0.78, and 1.11 mm) (Figure 3g-h,j). The α′ for the SAA crystal was calculated to be 0.0117 dB μm$^{-1}$ (Figure 3k).

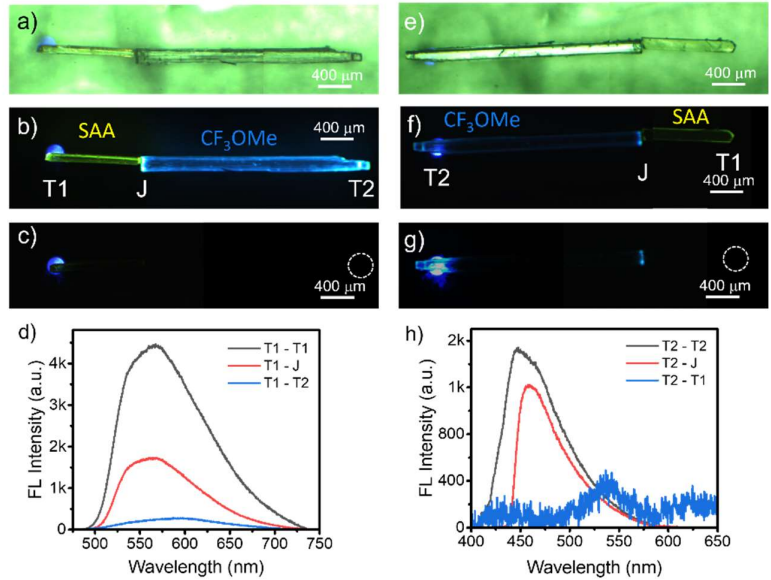

**Figure 4.** a) Confocal optical, b) FL (under UV torch excitation) and c) FL microscope (when LED light input is provided at crystal's left end) images of SAA-CF$_3$OMe photonic system. e) Confocal optical, and f,g) FL images of CF$_3$OMe-SAA photonic system when LED light input is provided at CF$_3$OMe tip. d-h) FL spectra recorded at T1 and T2 when LED light is provided at d) T1 and h) T2.

      The construction of photonic circuits relies on the rational use of active/passive waveguiding in the circuit design. Therefore, a hybrid active/passive waveguiding optical system using CF$_3$OMe and SAA crystal waveguides required the appropriate positioning of two crystals near the LED light source. Initially, both the crystals were transferred onto the same glass slide, followed by careful repositioning with the help of a sharp stainless-steel needle, resulting in the hybrid waveguide (the manipulation details are provided in SI; Figure S1). The SAA crystal and CF$_3$OMe are connected in a tip-to-tip fashion to ensure effective coupling from one waveguide to another (Figure 4a). The FL image of the hybrid system shows the bright body and tips of SAA and CF$_3$OMe crystals when excited with a UV torch (Figure 4b). When the hybrid optical waveguide was illuminated with UV LED light at the left end of the SAA crystal, labeled as T1 (Figure 4c) a bright yellow luminescence was produced. The same signal propagated through the SAA crystal to the junction, J (Figure 4b). At the junction, J, the yellow FL of the SAA crystal evanescently coupled into the CF$_3$OMe crystal and traveled to the other end of the CF$_3$OMe crystal at T2. As there is no energy transfer possibility from SAA to CF$_3$OMe, only the passive signal was outcoupled into CF$_3$OMe, and the same was recorded at T2 (Figure 4d).

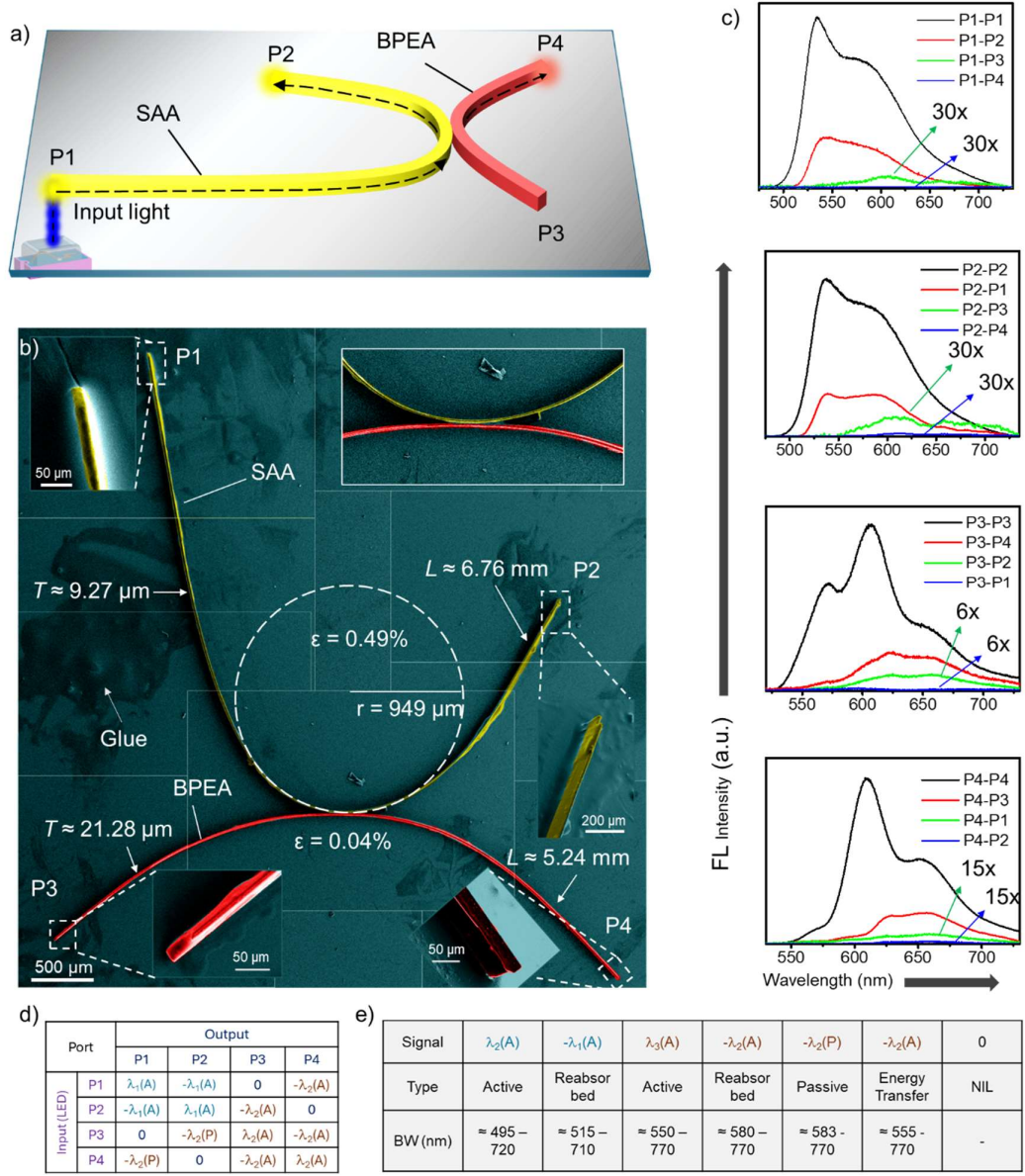

**Figure 5.** a) Graphical representation of a directional coupler of SAA and BPEA flexible crystals excited with LED at P1. b) Stitched color-coded FESEM images of the directional coupler. The inset shows the zoomed FE-SEM image of the junction of the waveguides. c) FL spectra at various outputs for input provided at P1, P2, P3 and P4 respectively. d) Input-dependent directional specific optical outputs obtained in the fabricated hybrid directional coupler. e) The table contains the signal type and their specific wavelength ranges.

Later, to illustrate active waveguiding from two hybrid crystal waveguides, a clever design strategy was adopted. The SAA and $CF_3OMe$ waveguides were repositioned with the help of a sharp needle to ensure effective optical signal coupling from earlier to the latter (Figure S2). Because of energy transfer from $CF_3OMe$ crystal to SAA crystal, with single input

LED light (λ ≈406 nm) to $CF_3OMe$, two outputs (blue $\lambda_1$ and yellow $\lambda_2$) are possible in the coupled photonic system. The optical performance of the modified hybrid optical system was studied by introducing a UV LED at the left end of the $CF_3OMe$ crystal. The dark field image shows bright blue emission at the point of excitation and the right end of CF3OMe crystal near the junction (Figure 4f-g). The FL spectra recorded at the excitation point (T2) and junction (J) reiterate the blue FL ($\lambda_1 \approx$ 420–670 nm) produced by $CF_3OMe$ crystal (Figure 4h). This blue FL emitted at the junction is sufficient to excite the SAA crystal due to energy transfer from $CF_3OMe$ to SAA thus, producing yellow FL ($\lambda_2 \approx$ 490–720 nm) at the left end of the SAA crystal at J, and the same propagates to the other end of the SAA crystal (T1), resulting in the active yellow signal propagation at the right end of the SAA crystal. Thereby, we could successfully demonstrate the LED-powered hybrid optical system functioning as active/passive waveguides operating with low-power LED light. Hybrid active-passive waveguiding of $CF_3OMe$-BPEA and BPEA-SAA crystal combination are shown in Figure S3.

For the construction of photonic circuits, steering light through different directions and pathways is crucial. Hence, it is customary to illustrate light propagation in bent waveguides, and their subsequent use in hybrid directional couplers (HDC) will prove useful. Therefore, a SAA crystal was bent using a sharp metallic needle into a curved geometry. This bent geometry was retained by fixing it with glue on a glass coverslip and placed on the device setup (Figure 5a). The bent SAA crystal also guided light similar to the nascent straight crystal under LED illumination with a $\alpha' \approx$ 0.003 dB $\mu m^{-1}$ (Figure 5e and S7a-j). In view of demonstrating a functional DC, bent SAA was carefully integrated with a BPEA waveguide ($L\approx$ 5.24 mm; $\alpha'$ =0.004 dB $\mu m^{-1}$) to create a HDC (Figure 5a,b, and S7k-s). The FESEM images reveal $T$ of 9.27 and 21.28 μm for the SAA and BPEA waveguides, respectively, and a clear physical contact between the two (Figure 5b). Various output ports in HDC are represented as P1-P4. The optical signals operational in HDC are yellow emission ($\lambda_2$(A) and -$\lambda_2$(A); here, minus and A represent reabsorbed and active type, respectively) from SAA and the orange signal ($\lambda_3$(A), -$\lambda_3$(A) and -$\lambda_3$(P); here P indicates passive light propagation) from BPEA waveguides.

The optical performance of the fabricated HDC was tested by providing optical excitation via the LED light at P1 and recording the transduced signals at other ports. At P1, a yellow signal $\lambda_2$(A) was produced, and the same was transduced to port P2 (Figure 5c,d). At P4, an orange signal due to ET from SAA to BPEA waveguide was evident. However, the restricted optical signal propagation in the backward direction prohibits any light seepage towards P3 and hence no FL (0) is recorded at P3 (Figure 5e). The optical input at P2 resulted in a yellow signal at P1 and an orange signal at P3, but no optical signal at P4 due to the directional light guiding ability of DC-2. Interestingly, when P3 or P4 receives the input light only one type of optical signal is operational in HDC because of no ET possibility from BPEA to SAA. Therefore, for P3 or P4 excitation, an orange signal is observed at P2 or P1 depending on the port receiving the input light. Finally, a novel approach to power organic crystalline waveguides, cavities or hybrid active/passive optical waveguiding systems for photonic applications was successfully demonstrated.

## 3. Conclusion

In summary, a novel design strategy was employed to create focused LED-based visible light sources for illuminating organic optical waveguides. Leveraging the fluorescence-based active waveguiding properties of blue ($CF_3OMe$), yellow (SAA), and orange (BPEA), a demonstration of UV LED-powered photonic systems was accomplished, spanning the entire visible spectral bandwidth from blue to near-infrared (NIR). This marks the first instance of focused LED-powered organic optical waveguide-cavities and hybrid active/passive waveguides. The experimental setup included the evaluation of a custom-designed hybrid directional coupler using flexible organic crystals (SAA and BPEA). These advancements hold the potential for groundbreaking developments in the realm of sustainable organic crystal-based visible light communications (VLCs) in the foreseeable future.

**Acknowledgements**

RC thanks SERB-New Delhi for the STAR Award (STR/2022/00011) and Core Research Grant (CRG/2023/003911). AK thanks PMRF for the fellowship.

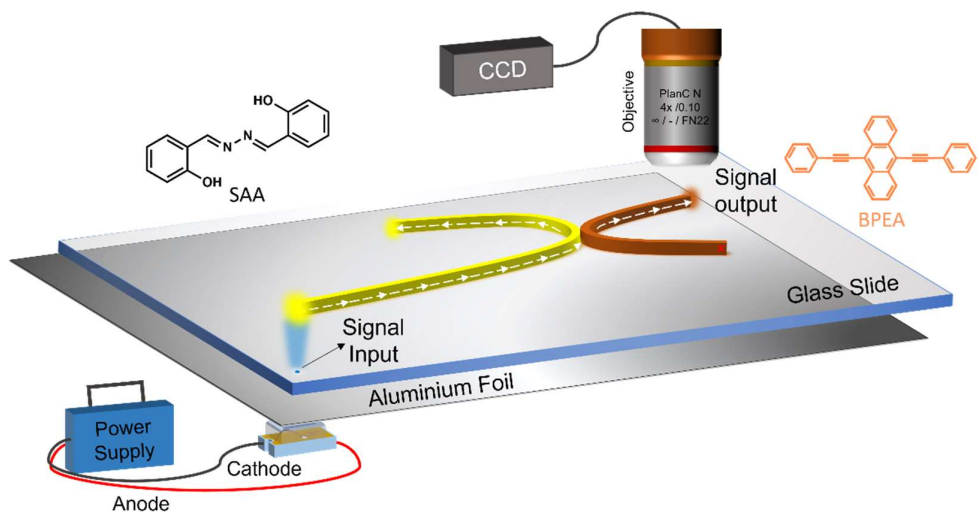

The usage of commercial LEDs for powering photonic devices is a starting point towards harnessing plenty of ambient light present around us. We demonstrate the LED-powered optical circuit device using flexible organic crystals for sustainable photonics.